\documentclass[prc,twocolumn,aps,showpacs,superscriptaddress]{revtex4}
\usepackage{amsmath}
\usepackage{amsfonts}
\usepackage{graphicx}
\begin{document}

\title{Effective non-Hermitian Hamiltonian and continuum shell model}

\author{Alexander Volya}
\affiliation{Physics Division, Argonne National Laboratory, Argonne, Illinois
60439}
\email{volya@anl.gov}
\author{Vladimir Zelevinsky}
\affiliation{National Superconducting Cyclotron Laboratory, Michigan State
University, East Lansing, Michigan 48824}
\affiliation{Department of Physics and Astronomy, Michigan State University,
East Lansing, Michigan 48824}
\begin{abstract}
The intrinsic dynamics of a system with open decay channels is described
by an effective non-Hermitian Hamiltonian which at the same time allows one to
find the external dynamics, - reaction cross sections. We discuss ways of
incorporating this approach into the shell model context. Several examples of
increasing complexity, from schematic models to realistic nuclear
calculations (chain of oxygen isotopes),
are presented. The approach is capable of describing a multitude of phenomena
in a unified way combining physics of structure and reactions.
Self-consistency of calculations and threshold energy dependence of the
coupling to the continuum are crucial for the
description of loosely bound states.

\end{abstract}
\date{\today}
\maketitle

\section{Introduction}

The center of interest in modern nuclear physics has recently moved toward
nuclei far from the valley of stability. Weakly bound nuclei cannot be fully
described in the limited framework of the shell model with a discrete energy
spectrum. Even the properties of their bound states reflect the proximity
of the continuum. Loosely bound nucleons create an extended spatial structure
that determines the results of possible reactions so that nearly all excitation
mechanisms break up the nucleus. The standard approaches of
many-body theory, such as the Hartree-Fock-Bogoliubov mean field and
random phase approximation, necessarily include virtual and real
excitations to the continuum. The Borromean cases of $^{6}$He, $^{9}$Be, and
$^{11}$Li, when the system can be considered to be made of three clusters with
all two-body subsystems being unbound, are very sensitive to the continuum
physics. This is the area where the conventional division of nuclear physics
into ``structure" and ``reactions" becomes inappropriate, and the two views
of the process, from the inside (structure and properties of bound states) and
from the outside (cross sections of reactions), should be recombined.

The broad success of the nuclear shell model with effective interactions
urges one to look for ways to incorporate the rich experience accumulated
in the shell model into a more general context which would properly include
the continuum part. We will not discuss below the most complicated task in
this direction, namely the problem of the effective interaction. It is
virtually unknown what should be an effective interaction of quasiparticles in
the restricted shell model space which includes the continuum. The work
that started with the Brueckner-Bethe theory of the $G$-matrix has to be
reviewed and modified.

Our goal here is much less ambitious. We would like to demonstrate the new
qualitative effects that emerge with the simple reformulation of the shell
model in terms of an effective non-Hermitian and energy-dependent Hamiltonian
describing the ``inside" view of the dynamics in a many-body system of
interacting particles coupled to and through the decay channels. For our
limited purpose below we assume that the effective
interaction of the shell model can be simply readjusted to the new problem,
although in fact it can be non-Hermitian by itself.

The description with the aid of an effective non-Hermitian Hamiltonian is well
known going back to the classical Weisskopf-Wigner damping theory \cite{weisskopf30},
works in atomic physics by Rice \cite{rice33} and Fano \cite{fano35} and
projection formalism by Feshbach \cite{feshbach58}.
The consistent formulation of the approach was given in the book by Mahaux and
Weidenm\"{u}ller \cite{mahaux69} in application to processes with
one particle in
the continuum. This gave rise to the shell model embedded in the continuum
\cite{barz77,rotter91} recently revived
\cite{bennaceur99,michel02,betan02} for the
description of loosely
bound nuclei. Another direction of development was related to the description
of statistical and chaotic phenomena in nuclear reactions
\cite{verbaarschot85,lewenkopf91}
and generalization of random matrix theory
\cite{sokolov88,fyodorov97} for unstable
systems. The detailed study of the effective non-Hermitian Hamiltonian revealed
new collective phenomena
\cite{sokolov88,sokolov92} with bright manifestations in
nuclear physics of low \cite{sokolov90,sokolov_rotter97}
and intermediate \cite{auerbach94,auerbach02}
energies, atomic physics \cite{magunov99,flambaum96},
molecular physics \cite{pavlov88}, quantum chemistry \cite{desouter99},
and
condensed matter physics \cite{seba00,nazmitdinov01,pichugin01}.
The basic origin of this collectivity
is the same as in the Dicke superradiance \cite{dicke54}, coherent coupling of
intrinsic states through common decay channels (common radiation field
of atoms confined to a small volume in the Dicke case). The ideas related to
this approach were used for an analysis of experimental data, especially in
two-state examples \cite{brentano96,sokolov94} taken from nuclear and
mesonic physics
as well as from the microwave cavity
experiments \cite{philipp00,persson00,stockmann02}.
Here we follow
a generic path of the shell model, adding continuum effects by including
explicitly non-Hermitian terms in the Hamiltonian
\cite{zelevinsky02,brentano02}.

\section{Non-Hermitian Hamiltonian}

We will not repeat here the full derivation of the effective
non-Hermitian Hamiltonian that can be achieved by separating the
full Hilbert space into the intrinsic part and the continua and
eliminating the continuum part with the aid of projection
operators. This procedure was addressed in detail by many authors,
see for example \cite{rotter91,rotter01}. We label intrinsic
states by {\sl 1,2,...}, and the continuum channels by
$a,b,c,...$. The matrix elements of the effective intrinsic
Hamiltonian can be written as
\begin{equation}
{\cal H}_{12}=H_{12}+\Delta_{12}-\frac{i}{2}W_{12},              \label{1}
\end{equation}
where $H$ is an internal, let us say, a standard shell-model part, and the last
two terms, which in general are functions of running total energy $E$, are
generated by the exclusion of the continuum.

The imaginary part $W(E)$ originates
from the real processes of decay to channels that are open at a given energy.
It is represented by the residues of the on-shell terms corresponding to the
delta-functions coming from the energy conservation and causality requirement
imposed on the energy denominators, $E\rightarrow E^{(+)}$. The quantity $W$
has a factorized form,
\begin{equation}
W_{12}=\sum_{c;{\rm open}}A_{1}^{c}A_{2}^{c\ast},                \label{2}
\end{equation}
where the decay amplitudes $A_{1}^c(E)$ are the matrix elements of the original
total Hermitian Hamiltonian between the states $|{\sl 1}\rangle$ and
$|c;E\rangle$ of different subspaces; the normalization coefficients are
included in the definition of $A_{1}^c$. The second term of Eq. (\ref{1}),
$\Delta_{12}(E)$,
originates from the principal value of the same expression and corresponds to
the virtual off-shell processes taking place via the continuum. Therefore it
includes contributions from all, open and closed, channels.  For the system
invariant under time reversal, one can use a real intrinsic basis, where
the matrix elements $H_{12}, \Delta_{12}$ and $A_{1}^c$ can be taken real.

The same effective Hamiltonian (\ref{1}) determines the scattering amplitude
and the reaction cross sections. The relation between the inside and outside
views was studied in \cite{mahaux69,sokolov88,sokolov92,sokolov90,sokolov_rotter97,sokolov97}.
The scattering matrix
in the channel space describing the $b\rightarrow a$ process is given by
\begin{equation}
S^{ab}=(s^{a})^{1/2}(\delta^{ab}-T^{ab})(s^{b})^{1/2}, \label {3+}
\end{equation}
\begin{equation}
T^{ab}=\sum_{12}A^{a\ast}_{1}\,\left(\frac{1}{E-{\cal H}}\right)_{12}
A^{b}_{2}.                                                        \label{3}
\end{equation}
Here $s^{a}=\exp(2i\delta_{a})$ stands for the smooth scattering
phase coming from remote resonances not accounted for explicitly.
The propagator $(E-{\cal H})^{-1}$ in the scattering amplitude
$T^{ab}$ does not depend on a specific reaction and contains the
full effective Hamiltonian (\ref{1}) with the same amplitudes
$A_{1}^{c}$ as those determining the entrance and exit channels in
Eq. (\ref{3}). This guarantees the unitarity of the $S$-matrix
since the virtual processes of evolution of the open system to and
from the continuum channels are included in all orders in the
propagator. Indeed, if one introduces the intrinsic Hermitian
propagator $(E-H)^{-1}$ and the second order Hermitian scattering
amplitude,
\begin{equation}
K^{ab}=\sum_{12}A^{a\ast}_{1}\,\left(\frac{1}{E-H}\right)_{12}A^{b}_{2},
                                                              \label{3a}
\end{equation}
where the propagation does not include the coupling to continuum, the full
scattering amplitude is given by the geometric series (the hats mark the
operators in the channel space),
\begin{equation}
\hat{T}=\frac{\hat{K}}{1+(i/2)\hat{K}}.                     \label{3b}
\end{equation}
Then the scattering matrix is explicitly unitary,
\begin{equation}
\hat{S}=\hat{s}^{1/2}\frac{1-(i/2)\hat{K}}{1+(i/2)\hat{K}}\,\hat{s}^{1/2}.
                                                                \label{3c}
\end{equation}

The diagonalization of the non-Hermitian Hamiltonian (\ref{1}) produces the
complex eigenvalues
\begin{equation}
{\cal E}_{\alpha}(E)=\tilde{E}_{\alpha}(E)-\frac{i}{2}\Gamma_{\alpha}(E),
                                                                     \label{4}
\end{equation}
where the real, $\tilde{E}_{\alpha}$, and imaginary, $\Gamma_{\alpha}$, parts
are functions of running real energy $E$. Without explicit energy dependence
of the effective Hamiltonian, these eigenvalues would provide the unstable
states with a pure exponential decay law $\propto \exp(-\Gamma_{\alpha}t)$.
The presence of energy dependence violates the exponential decay, and the
actual quasistationary states are found at real energies $E_{\alpha}$
determined by the self-consistency condition
\begin{equation}
\tilde{E}_{\alpha}(E_{\alpha})=E_{\alpha}.                      \label{5}
\end{equation}
The line-shape is not Breit-Wigner but we still call $\Gamma_{\alpha}
(E_{\alpha})$ the width of the resonance $\alpha$. In what follows we
omit the tilde sign for $E_{\alpha}$ if it does not lead to a confusion.

In the region of
interest, namely for loosely bound systems, the main energy dependence comes
from the proximity of thresholds as was stressed in Refs. \cite{sokolov92,brentano02}.
The channel $c$ is open only if the total energy $E$ is above the
threshold energy
$E^{(c)}$ for this channel. The decay amplitudes associated with the channel
$c$ contain therefore the step factor $\Theta(E-E^{(c)})$ and can be written as
\begin{equation}
A_{1}^{c}=a_{1}^{c}(E)\Theta(E-E^{(c)}),                       \label{6}
\end{equation}
where $a_{1}^{c}(E)$ is a smooth function of energy that falls off to zero when
energy decreases to the threshold value. For a single-particle
decay channel, it can be parameterized \cite{sokolov92,brentano02} as proportional to the
square root of the penetrability in this channel.

The real part $\Delta$ of
the effective potential can be written as the principal value integral
\begin{equation}
\Delta_{12}(E)=\frac{{\cal P}}{\pi}\sum_{c}\int_{E^{(c)}} \frac{dE'}{E-E'}
a_{1}^{c}(E')a_{2}^{c\ast}(E').                         \label{7}
\end{equation}
Under the same assumption of a non-singular character of $a_{1}^c$, the matrix
elements (\ref{7}) also have a smooth energy dependence with no singularities
near threshold, and can be approximated by energy-independent quantities as
in \cite{sokolov92,brentano02}.

One formal conclusion concerning the existence of bound and unbound states can
be reached just from the way the theory is constructed. If the conventional
shell model with a purely discrete spectrum (no coupling to the continuum)
predicts a state with energy below all decay thresholds, this state will
remain bound in the full calculation with the decay amplitudes included.
Indeed, all widths depend on the total energy and vanish
below thresholds so that
the old solution is still valid. However, this statement is formal since it
assumes that the reaction thresholds are known beforehand. In fact, they have
to be determined consistently for the chain of nuclides relevant to the
reactions under consideration.

Certainly, there are limitations in the applicability of the
effective non-Hermitian Hamiltonian method in the form outlined in
the present article. As energy increases, a rapid growth of a
number of interfering open channels makes this approach
impractical. The approximations of a different type can be then
introduced in the general framework, see for example
\cite{sokolov97,zelevinsky98}, that directly lead to quantum kinetics of
statistical reactions \cite{bortignon91}. We also deliberately limit
ourselves here by taking into account only the energy-dependence
associated with threshold and resonance phenomena although the
smooth ``potential" scattering part could be included without
significant difficulties via the entrance and exit scattering
phases hidden in the factors $s^{a}$ and $s^{b}$ of Eq.
(\ref{3+}). The full energy dependence was discussed, in
particular, in Refs. \cite{stockmann02,drozdz00}. The main
physical assumption made here is that the states under
consideration are close to threshold and, at relatively low
energy, only few open channels are really essential. The
deviations resulting from violations of these conditions have been
studied numerically in \cite{pichugin01}. For the purpose of this
paper, namely for the development of shell model methods intended
for the description of low-lying states in the nuclear systems
near the border of stability, the presumed conditions are usually
fulfilled to within a sufficient accuracy.

\section{Shell model approximation}

We view the Hamiltonian ({\ref{1}) as a sum of three terms,
\begin{equation}
{\cal H}=H^{\circ}+V-\frac{i}{2}\,W\,,
\end{equation}
where we assume that the intrinsic Hermitian part $H^{\circ}+V$
consists of independent particle energies,
\begin{equation}
H^{\circ}=\sum\epsilon_{\nu}a^{\dagger}_{\nu}a_{\nu},          \label{8}
\end{equation}
and the effective Hermitian interaction $V$. As a renormalization
of the standard shell model interaction, the Hermitian matrix
elements $\Delta_{12}$, Eq. (\ref{7}), generated by the virtual
coupling through continuum, can be incorporated into the operator
$V$. The approximation of energy independence of the operator
$\Delta\,$ and, as a result, energy independence of $V\,,$ also
used in previous works \cite{sokolov92,brentano02}, can be easily
removed.

In order to formulate physical problems in the spirit close to the
conventional shell model, we start from the basis states
$|\Phi\rangle\,,$ the eigenstates of $H^{\circ},$ that are Slater
determinants of the $m$-scheme or their linear combinations
projected onto correct values of total spin $J$ and isospin $T$.
For the next step we consider the ``unperturbed" part of the
Hamiltonian which includes the independent particle part
$H^{\circ}$ and the imaginary part $-(i/2)W$
\begin{equation}
{\cal H}^{\circ}=H^{\circ}-\frac{i}{2}W.                      \label{11}
\end{equation}
The diagonalization of the Hamiltonian (\ref{11}) along with the
self-consistent solution of Eq. (\ref{5}) gives new eigenvectors
$|\tilde{\Phi}\rangle$ either with complex energies (\ref{4}) or
as stable configurations on the real energy axis. The factorized
nature of the operator $W$, Eq. (\ref{2}), that is preserved by
orthogonal transformations, and the presence of symmetries may
bring additional simplifications. In some special cases,
 ${\cal H}^{\circ}$ remains diagonal in the original basis
$|\Phi\rangle\,.$ In these situations the meaning of the
amplitudes is the most clear, being related to the single-particle
decay into continuum.

Thus, for an isolated single-particle level $|\nu)$ embedded in
the continuum, the unperturbed real energy is $E_{{\rm core}}
+\epsilon_{\nu}$. If the only open channel, $c\Rightarrow \nu$, is
associated with the emission of the particle $\nu$, the imaginary
part $W$ leads to the width
\begin{equation}
\gamma_{\nu}=|A_{\nu}^{\nu}|^{2}                              \label{9}
\end{equation}
for any configuration which consists of the particle on the level
$|\nu)$ and an arbitrary state of the stable core (no interaction
between them at this stage). Threshold energy is determined by the
core configuration. Similarly, in the case of several
single-particle levels $\nu$ embedded in the continuum, the
single-particle decay channels, opened for a specific
configuration $|\Phi\rangle$ with occupation numbers
$n_{\nu}(\Phi)=0$ or 1, result in the width
\begin{equation}
\gamma(\Phi)=\sum_{\nu}n_{\nu}(\Phi)\gamma_{\nu}.           \label{10}
\end{equation}

If there are several single-particle levels $|\nu_{i})$ with the
same exact quantum numbers $j^{\pi}\tau$, the situation is more
complicated and in general $H^{\circ}$ and ${\cal H}^{\circ}$
cannot be simultaneously diagonalized. Apart from the particle
emission from a given single-particle state we have now also the
interaction through continuum given by the off-diagonal elements
of $W_{12}$. Here the Dicke collectivization and redistribution of
the widths \cite{sokolov88,rotter91,sokolov92} are
possible, see below, and already at this stage, with no residual
shell-model interaction $V$, we need to diagonalize the
non-Hermitian Hamiltonian ${\cal H}^{\circ}\,.$

This situation is almost certainly present in cases where
two-particle emission is possible from different initial
configurations. For example, a zero spin pair can be emitted from
a few $j$-levels leading to a final state with the same quantum
numbers and therefore into the same decay channel. Here the
coupling through the continuum may be very important.

In the most general case, because of the energy dependence in the
amplitudes $A_{1}^{c}$, the positions of the resulting
quasistationary eigenstates of ${\cal H}^{\circ}$ in the complex
plane should be determined avoiding false solutions that can
appear due to the possible non-analytic energy dependence at
thresholds. One can take only those complex poles that can be
traced back to the real axis (independent particle states
$|\Phi\rangle$) in the case of the closed channels. Another new
feature is that, generally speaking, the threshold energies are
not known {\sl a-priori}. They are to be calculated
self-consistently comparing total energies of the parent and
daughter nuclei taken in the same approximation. But this must be
done only on the next step when the residual shell-model
interaction is accounted for.

Finally, we include an effective interaction $V$ that introduces
the mixing of (in general unstable) shell model configurations
$|\tilde{\Phi}\rangle$. As a result, instead of original
independent particle states, we obtain the states $|\Psi\rangle$
which characterize both the intrinsic structure and possible decay
channels in the fully interacting system. Since the Hamiltonian is
non-Hermitian, the resonance ``energies" (\ref{4}) move in the
complex plane relative to their initial positions given by
eigenvalues of ${\cal H}^{\circ}$ (states $|\tilde{\Phi}\rangle$)
and this dynamics may be quite complicated driving some states
back to stability.

Another important ``intermediate'' Hamiltonian $H^{\circ}+V$
describes the case when all decay widths are ``switched off'' and
obviously corresponds to the standard shell model. Below we denote
shell-model many-body eigenstates as $|\Psi_{{\rm s.m.}}\rangle$.
Of course, in practice it is not necessary to make a two-step
diagonalization, and the intermediate steps with the wave
functions $|\tilde{\Phi}\rangle$ or $|\Psi_{{\rm s.m.}}\rangle$
can be avoided.

It is known, see for instance \cite{baz71}, that the eigenstates
of a non-Hermitian Hamiltonian form a biorthogonal system which
would allow one to study the observable characteristics of
unstable states along with the reaction cross sections and
dynamics transformed to the time domain. Below we show several
examples ordered by increase of complexity, from very schematic to
more realistic. The selected cases illustrate the diversity of the
physical phenomena that can appear in unstable many-body systems
and can be described by the method of an effective non-Hermitian
Hamiltonian. As mentioned in the Introduction, in this exploratory
study we assume the effective interaction to be known and its
matrix elements $V_{12}$ (taken in a basis of stable states and
including $\Delta_{12}$) to be real.

\section{Single-particle decay in a many-body system}
\label{sec:subspace}

\subsection{One single-particle level in the continuum;
energy-independent continuum width}

We start with the simplest problem (a similar example was shown in Ref.
\cite{zelevinsky02}). Consider a set of single-particle energies
$\epsilon_{\nu}$ with the upper of them lying above the particle emission
threshold $\epsilon^{(c)}$, see Fig. \ref{decscheme},
where a system of $\Omega$
single-particle levels is presented with $\epsilon_{\nu}<\epsilon^{(c)}$
for $\nu =1,...,\Omega-1$
and $\epsilon_{\bf \nu}>\epsilon^{(c)}$ for ${\bf \nu}=\Omega$.
For simplicity we assume here that the levels are
equidistant on the real axis and not degenerate; later we add the Kramers
double degeneracy. We assume that all single-particle emission channels have
different quantum numbers and cannot be coupled through continuum. Thus, we
have only one initial non-zero single-particle width expressed with the aid of
the complex single-particle energy
\begin{equation}
e_{\nu}=\epsilon_{\nu}-\frac{i}{2}\gamma\,\delta_{\nu \Omega}.     \label{12}
\end{equation}
Now we use this set of single-particle levels as the basis for forming the
many-body configurations $|\Phi\rangle$ as Slater determinants with all
possible distributions of the occupancies $n_{\nu}$. Finally we switch on a
real two-body interaction $V$.
\begin{figure}
\begin{center}
\includegraphics[width=5 cm]{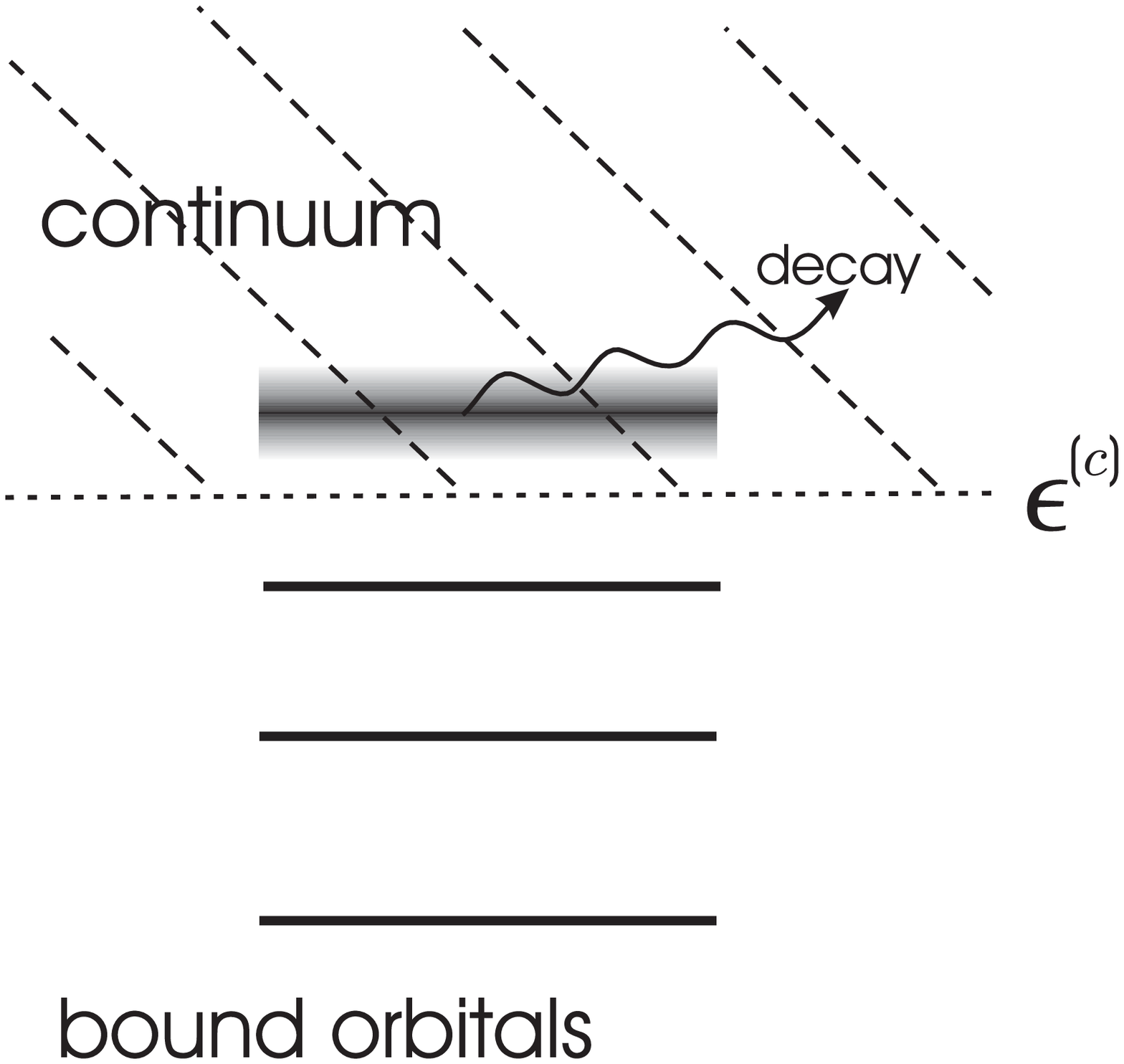}
\end{center}
\caption{\label{decscheme}}
\end{figure}

In order to characterize the generic results which are insensitive
to specific peculiarities of the residual interaction, in this
example we use a system of $N=4$ fermions with the equidistant
spectrum $\epsilon_{\nu}$ of $\Omega=8$ orbitals and random
(Gaussian distributed) matrix elements of the two-body
interaction. We solve this problem by diagonalizing the complex
matrix of the full many-body Hamiltonian ${\cal H}$. The results
are shown in Fig. \ref{wdn4l8_f1} as the dynamics of the complex
eigenvalues evolving as a function of the only variable parameter,
the single-particle width $\gamma$, taken here as an
energy-independent number.

At $\gamma=0$ (a normal shell-model limit), the many-body states
$|\Psi_{{\rm s.m.}}\rangle$ obtained with the real residual
interaction are stable, and their spectrum can be represented by
the points on the real axis. As $\gamma$ increases, all states
$|\Psi_{{\rm s.m.}}\rangle\Rightarrow |\Psi\rangle$ acquire widths
and move into the complex plane. This means that, because of the
unrestricted configuration mixing, any many-body eigenstate
$|\Psi\rangle$ contains an admixture of configurations with the
occupied upper orbital, and therefore it is capable of decay. In
the limit of small $\gamma\,,$ in agreement with usual
perturbation theory it is expected that $|\Psi\rangle\approx
|\Psi_{{\rm s.m.}} \rangle\,.$ Thus, the decay width of a
many-body state $\Psi$ is, similarly to the case of Eq.
(\ref{10}), determined by the spectroscopic factor of a progenitor
stable state $|\Psi_{{\rm s.m.}} \rangle$ that is given as an
occupation probability of a decaying single-particle orbital ${\bf
\nu}$,
\begin{equation}
\Gamma(\Psi)=\gamma n_{\bf \nu}(\Psi_{{\rm s.m.}})\,.       \label{12a}
\end{equation}
This natural picture breaks down once the value of $\gamma$
becomes comparable to the level spacing $D$ along the real axis,
and the internal dynamics gets affected by the continuum in a
non-perturbative way.

In this example $\gamma$, and therefore the effective Hamiltonian
${\cal H}$, is independent of running energy $E\,$. This makes the
trace of ${\cal H}$ a conserved quantity resulting in
\begin{equation}
-2 \,{\rm Im}({\rm Tr}\,{\cal H})=\sum_{\Psi} \Gamma(\Psi) =
\frac{(\Omega-1)!}{(\Omega-N)!\,(N-1)!}\,\gamma \,.
                                                   \label{sumgamma}
\end{equation}
The real part of the trace $\sum_{\Psi} \tilde{E}(\Psi)$ also
remains constant and is independent of $\gamma\,.$ In Eq.
(\ref{sumgamma}) we counted the number of non-interacting
configurations available for $(N-1)$ particles if the decaying
orbital is occupied. Conservation of the trace of ${\cal H}^2$
and the fact that the imaginary part $W$ is diagonal make
$(\gamma)^{-1}\,\sum_{\Psi} \tilde{E}(\Psi) \Gamma(\Psi)$ also a
$\gamma$-independent constant.

In the $\gamma\rightarrow 0$ limit, the occupancy of the decaying
orbital governs the distribution of widths in Eq.
(\ref{sumgamma}). This can be generalized by introducing the
dynamic occupation numbers
\begin{equation}
n_{\Omega}(\Psi;\gamma)
=\frac{\partial \Gamma(\Psi;\gamma)}{\partial \gamma}\,.   \label{dynocc}
\end{equation}
These parameters describe how at a given $\gamma$ an infinitesimal
increase of the initial single-particle width $\gamma$ is
distributed among the many-body states $\Psi$. According to Eq.
(\ref{sumgamma}), $\sum_\Psi  n_\Omega(\Psi, \gamma)$ is
independent of $\gamma\,.$ Despite all the resemblance to
occupation numbers the numbers (\ref{dynocc}) can be negative. One
can also introduce generalized spectroscopic factors
$\Gamma(\Psi;\gamma)/\gamma$ that are always positive and bound
between 0 and 1, but being cumulative quantities they would be
less sensitive to dynamical features.

With further increase of $\gamma$ the picture in Fig.
\ref{wdn4l8_f1} looks paradoxical \cite{zelevinsky02}. The
many-body states are clearly divided into two groups. The complex
energies of the first group rapidly move away from the real axis
revealing large widths. At the same time the states of the second
group turn back to the real axis and keep only tiny widths, whence
becoming long-lived. Similar phenomena are known for a long time
from various schematic studies \cite{haake94,persson99} and
versions of the shell model with the continuum effects included,
for example \cite{moldauer75,kleinwachter85}.

The puzzle is readily resolved since with $\gamma$ increasing we
come to the situation where the imaginary term dominates the
dynamics and therefore classifies the eigenstates by their
relation to decay rather than by real energy. Any superposition of
configurations with a considerable amplitude of the occupied
unstable level {\sl 8} undergoes fast decay. Such states
constitute the first group. The superpositions of the
configurations with the empty level {\sl 8} become eigenstates of
the second group and correspond to long-lived compound states. It
is easy to calculate the dimensions of the two groups. The total
number of states ${\cal N}$ for 4 fermions on 8 non-degenerate
orbitals is $8!/(4!)^{2}=70$. The first group contains the states
where the level {\sl 8} is occupied and the remaining 3 particles
are distributed over 7 stable levels; the corresponding dimension
is $7!/(3!4!)=35$, so that the eigenstates are divided in this
case evenly between the two groups, in agreement with Fig.
\ref{wdn4l8_f1}.

We see that the strong coupling of intrinsic states to the
continuum produces a natural segregation
\cite{sokolov88} of processes into fast direct reactions
and slow compound nucleus reactions. Here the direct processes are
of single-particle nature, and at large $\gamma$ the corresponding
width of each state of the first group is close to $\gamma$. Fig.
\ref{wo}, where the segregation of the effective occupation
numbers is shown as a function of $\gamma$, confirms the division
of states into short-lived with a fully occupied decaying
single-particle orbital, $n_8=1$, and compound ones where
$n_8=0\,.$ The analog of this segregation effect has been
experimentally seen \cite{persson00} in microwave cavities,
although for realistic systems the situation is more complicated,
mainly because of the uncertainty in the effective number of open
channels \cite{drozdz00,nazmitdinov01}.

In order to further quantify this phase transition we introduce a
parameter $\xi$ which roughly shows the fraction of segregated
states lying in a small vicinity $\sigma$ near $n_8=0$ and
$n_8=1\,,$
\begin{equation}
\xi(\gamma)=\frac{1}{\cal N} \sum_{\Psi} \left [
e^{-n^{2}_\Omega(\Psi)/(2 \sigma^2)}+e^{-[1-n^{2}_\Omega(\Psi)]^2/
(2 \sigma^2)} \right ].                                 \label{segre}
\end{equation}
We select here $\sigma=0.1\,.$ In Fig. \ref{pt} the quantity $\xi$
is plotted as a function of $\gamma$ for various relative
strengths of residual mixing, panel ({\sl a}), and for different
half-occupied systems, panel ({\sl b}). It follows from the graphs
of Fig. \ref{wo} and \ref{pt} that the segregation starts at
$\gamma \sim D \,,$ and occurs gradually as $\gamma$ dominates the
residual mixing $V\,,$ resulting in a  peculiar phase transition.
The theory of this general phenomenon seen earlier in numerical
simulations \cite{moldauer75,kleinwachter85} was developed in
\cite{sokolov88} where an analogy to the Dicke coherent
state in optics \cite{dicke54} was pointed out. A similar effect,
with possible implications for quantum computers, is known in the
context of quantum measurement theory, where coupling and decay
created by the measurement mechanism can result in a separation of
dynamics into decoupled subspaces \cite{facchi02}.

Formally the phase transition follows from the factorized
structure of the continuum coupling in Eq. (\ref{2}). The rank $r$
of the matrix $W$ is equal to the number of open channels (the
dimension of the first group of states in the example of Fig.
\ref{wdn4l8_f1}) that is smaller than the total dimension ${\cal
N}$ of the intrinsic space. This matrix has $r$ nonzero
eigenvalues whereas the remaining ${\cal N}-r$ eigenvalues are
equal to zero. In a more specific language
\cite{sokolov88,sokolov92}, the independent particle
basis $|\tilde{\Phi}\rangle$ of this example is the ``doorway"
basis for the coupling to and through the continuum. In the limit
of strong continuum coupling, the part $W$ dominates the dynamics
and aligns the eigenstates along its eigenvectors. Similarly to
the coupling of two-level atoms through their common radiation
field in the Dicke superradiance, here the intrinsic states are
coupled through the decay channels. At values of $\gamma$
exceeding the level spacing (the regime of overlapping
resonances), this coupling becomes strong, and the continuum
accomplishes the self-organization of intrinsic structure
\cite{rotter91}. The real part $V$ of the interaction is needed
only to establish the correct level density on the real axis. The
random character of $V$ in the above example does not prevent the
system from the ordering by the continuum.
\begin{figure}
\begin{center}
\includegraphics[width=8 cm]{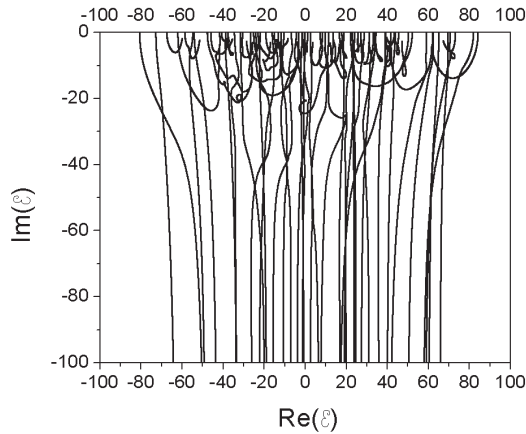}
\end{center}
\caption{Trajectories of 70 many-body states of the system of 4
particles on 8 single-particle levels are shown in a complex plane
as a function of the increasing decay width $\gamma$ of the upper
single-particle state. The residual two-body interaction matrix
elements are selected randomly from the Gaussian distribution with
zero mean and with the variance of one energy unit. The
single-particle energies are equidistant with a spacing
$\Delta\epsilon= 0.5$ energy units. \label{wdn4l8_f1}}
\end{figure}

\begin{figure}
\begin{center}
\includegraphics[width=8 cm]{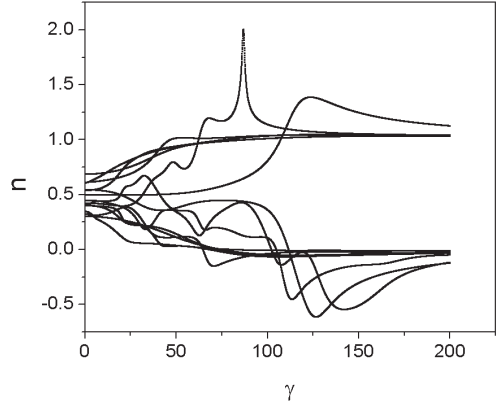}
\end{center}
\caption{The evolution of the effective occupation number $n_8$, Eq. (21),
as a function of $\gamma$ for the 13 lowest (selected at
$\gamma=0$) many-body states.
\label{wo}}
\end{figure}

\begin{figure}
\begin{center}
\includegraphics[width=8 cm]{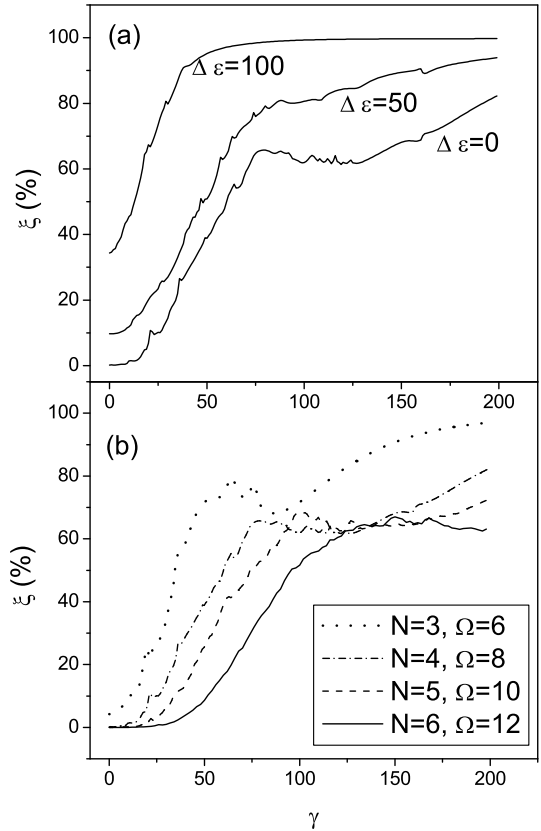}
\end{center}
\caption{Fraction of segregated states as a function of $\gamma\,$
for the same system as in Figs. 3 and 4. For the upper panel ({\sl
a}) the single-particle level spacing is varied from the
degenerate case, $\Delta \epsilon=0$, to $\Delta \epsilon =100$, a
point where the residual interaction $V$ can be completely
ignored. The lower panel ({\sl b}) shows the ``condensed''
fraction of many-body states for a degenerate single-particle
spectrum and various system sizes. In all cases only one
single-particle level undergoes decay with the width $\gamma\,$.
\label{pt}}
\end{figure}

\subsection{Two unstable single-particle levels}

A slightly more complex example is shown in Fig.  \ref{wdn4l8_f2}.
Here we again have eight equidistant non-degenerate
single-particle orbitals $\epsilon_{\nu}$. Two of them, with
different quantum numbers, have nonzero widths equal to $\gamma$
and $\gamma/2$. With the random interaction turned on, the
evolution of the 70 eigenvalues with increasing $\gamma$ separates
four groups of the eigenstates with different decay rates, fast,
slow and two intermediate. Short-lived states, type ({\sl a}) in
Fig. \ref{wdn4l8_f2}, include configurations with both unstable
orbitals being occupied, thus at large $\gamma$ their width is
roughly $\Gamma^{(a)}\approx 3 \gamma/2$. The number of such
configurations, $6!/(2!4!)=15$, is equal to the number of
long-lived states [type ({\sl d})]. The corresponding quasi-stable
configurations with $\Gamma^{(d)}\approx 0$ at $\gamma\rightarrow
\infty$ do not have a noticeable admixture of unstable orbitals so
that all four particles are distributed over six stable orbitals.
Finally, the intermediate lifetimes $\Gamma^{(b)}\approx  \gamma$
and $\Gamma^{(c)}\approx  \gamma/2$ correspond to cases where only
one of the unstable orbitals is filled. The number of such cases
is $6!/(3!)^{2}=20\,.$
\begin{figure}
\begin{center}
\includegraphics[width=8 cm]{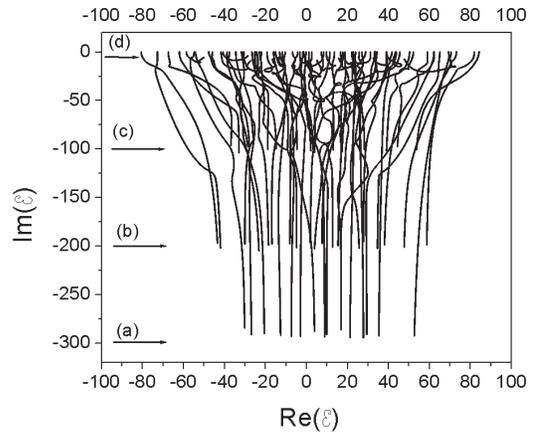}
\end{center}
\caption{\label{wdn4l8_f2}Trajectories of 70 many-body states of
the system of 4 particles on 8 single-particle levels are shown in
a complex plane as a function of the increasing decay width,
$\gamma$ and $\gamma/2$, of the two upper single-particle states.
The single-particle energies and the two-body interaction matrix
elements are the same as in Fig. 2. Arrows ({\sl a}), ({\sl b})
and ({\sl c}) indicate the endpoints at large $\gamma$ for the
corresponding types of states: short-lived ({\sl a}), and two
intermediate types ({\sl b}) and ({\sl c}), see text. The
remaining states of the fourth class ({\sl d}) are quasistable.}
\end{figure}

\subsection{Kramers degeneracy}
\begin{figure}
\begin{center}
\includegraphics[width=8 cm]{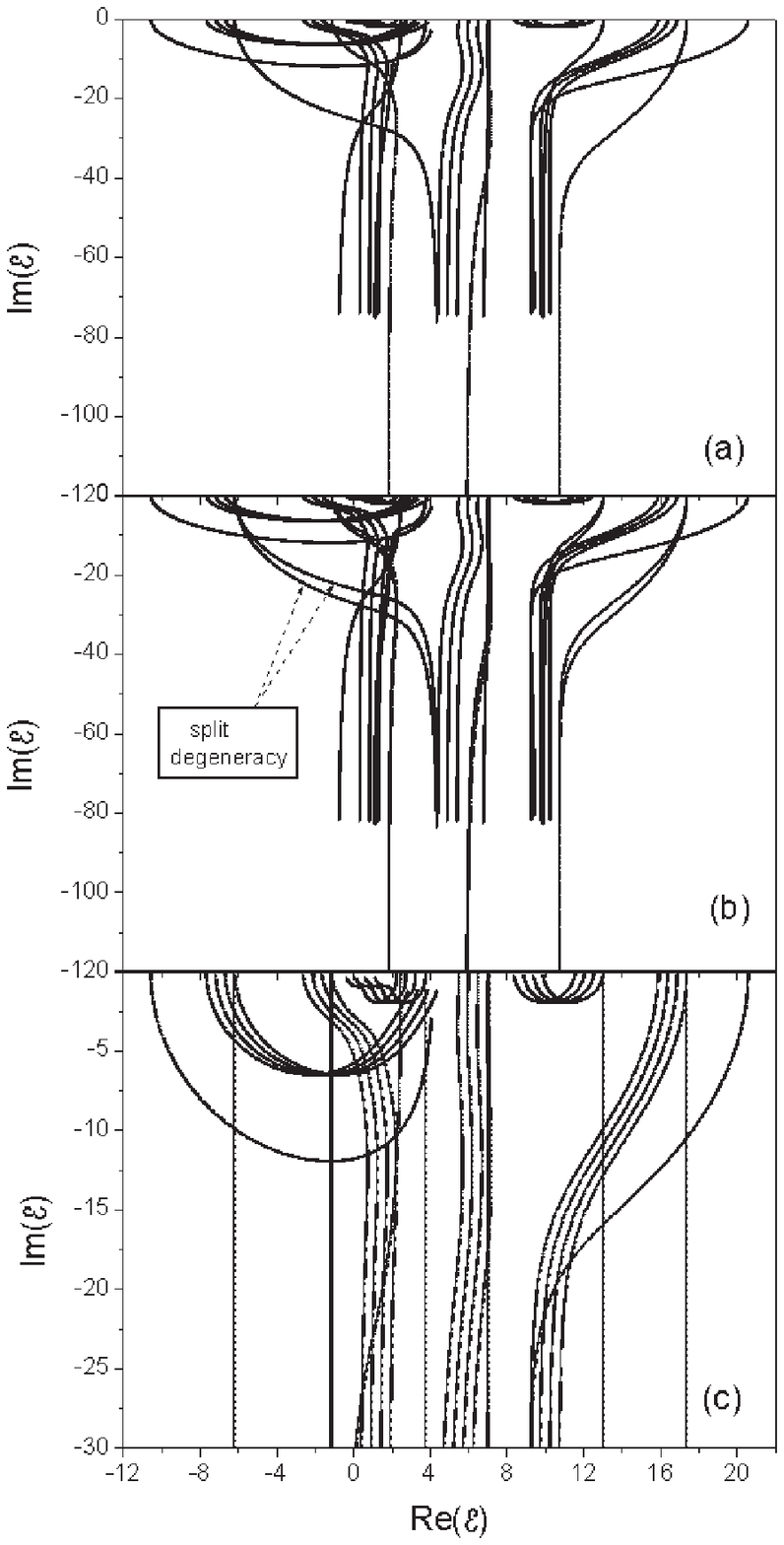}
\end{center}
\caption{Dynamics of many-body eigenstates of the system
containing 4 double-degenerate single-particle states and $N=3$
particles. Panel ({\sl a}) corresponds to the upper two degenerate
single-particle levels having the same width $\gamma\,.$ The
Kramers degeneracy is preserved in this case. Panels ({\sl b}) and
({\sl c}) correspond to the situations where the invariance is
broken by the decay. In ({\sl b}) the upper pair of initially
degenerate time-conjugate levels have widths $\gamma$ and $1.1
\gamma$, respectively. The case ({\sl c}) presents the maximum
symmetry violation as the time-conjugated levels have widths
$\gamma$ and 0. \label{kramers}}
\end{figure}

The manifestations of symmetries and symmetry breaking in open
systems are particularly interesting and important for the shell
model formalism. In the presence of global symmetry, such as
rotational invariance, the entire space that includes both
internal and external states, is symmetric. In this case
degeneracies of states are possible and there is no mixing between
the classes of states with different exact quantum numbers. A
non-trivial situation occurs when symmetry is violated in external
space or in the internal-external coupling, while the intrinsic
system is still symmetric. Then the degeneracy is lifted and in
the effective Hamiltonian, in general, both non-Hermitian part $W$
and Hermitian part $\Delta$ are no longer invariant. Here we
consider a schematic case when symmetry is violated only in the
internal-external coupling.

In systems invariant under time reversal, the states of an odd
number of fermions are at least double degenerate (Kramers
degeneracy). Let our eight single-particle orbitals form four
double-degenerate pairs, as it happens in the body-fixed frame of
a nucleus with static multipole deformation where the degeneracy
in an axially-symmetric case is connected with the sign of the
magnetic quantum number, $\pm m$. This situation is shown in Fig.
\ref{kramers}({\sl a}) for the case of 3 particles and random
selection of time-reversal invariant two-body residual
interaction. All curves in this figure correspond to the evolution
of double-degenerate many-body states, since two decaying
Kramers-degenerate single-particle levels are assumed to have the
same width, and thus the symmetry is not violated by decay.

The case ({\sl b}) illustrates the situation when time-reversal
invariance is slightly distorted in the decay channel, for
instance by an external magnetic field, so that one of the
Kramers-degenerate levels has a 10\% larger width. The long-lived
and short-lived states are the least affected ones, in Fig.
\ref{kramers}({\sl b}) the splitting of their degeneracy is hard
to resolve. Indeed, for both of these cases the decay dynamics
either lock a particle pair on the decaying time-conjugate
orbitals or make them both unoccupied. This restriction of motion
allows the system to retain the quasi-invariance. The remaining
states with the intermediate lifetime involve superpositions with
one particle being on either of the two decaying single-particle
states. Such a superposition, generally, is no longer
time-reversal invariant and the Kramers degeneracy is broken.
Finally, in the limit of strong decay, residual two-body
interactions become less effective in mixing and then real parts
of complex energies ${\cal E}$ regain the degeneracy. These
arguments, however, no longer hold true in the extreme limit of
the violated time-reversal invariance when one of the
time-conjugate orbitals becomes stable while the partner decays,
Fig. \ref{kramers}({\sl c}). Then the Kramers degeneracy is broken
and, since there is no intermediate groups, the states join
long-lived or short-lived families, which leads to
35(long)+21(short)=56 states.

\section{Dynamics of two states coupled to a common decay channel}

\subsection{Appearing of binding}
\label{sec:5A}

Here we consider a model that shows how the attractive real
interaction works generating the binding of originally unstable
states in the presence of the coupling through continuum. We
consider two single-particle levels, let say $s_{1/2}$ and
$p_{1/2}$ orbitals, in the continuum so that their energies
$\epsilon(s)$ and $\epsilon(p)$ are positive if the continuum
threshold is put at zero energy. In the three-body Borromean model
for $^{11}$Li with the inert core of $^{9}$Li and
particle-unstable $^{10}$Li, the two active states are those for a
pair of halo neutrons, $\epsilon_{1}=2\epsilon(p)$,
$\epsilon_{2}=2\epsilon(s)$. They are quasistationary, and their
decay amplitudes $A_{1,2}$ for the only open channel,
characterized by the core nucleus in the ground state and the
neutron pair in the state $J^{\pi}=0^{+}$ in the continuum, can be
found from a single-particle picture. At this point the exact form
of the energy dependence is not fixed, except for the fact that
when the total energy approaches zero, the decay becomes forbidden
so that, as in Eq. (\ref{6}), the amplitudes $A_{1,2}$ contain the
step function $\Theta(E)$. A special case with one initial
non-zero width was presented in \cite{brentano02}. The problem of
two-body decay, especially relevant for Borromean systems, was
recently approached with the use of different shell model
formalisms in Refs. \cite{michel02,betan02}. Two-proton
radioactivity \cite{pfutzner02} is another example requiring a similar
consideration.

According to Sect. II, the effective non-Hermitian Hamiltonian in this
$2\times 2$ space is
\begin{equation}
{\cal H}=\left(\begin{array}{cc}
\epsilon_{1}-\frac{i}{2}\gamma_{1}&v-\frac{i}{2}A_{1}A_{2}\\
v-\frac{i}{2}A_{1}A_{2}&\epsilon_{2}-\frac{i}{2}\gamma_{2}
\end{array}\right).                                      \label{13}
\end{equation}
Here $V_{12}\equiv v$ is the real mixing matrix element,
$\gamma_{1,2}= A_{1,2}^{2}$, and the amplitudes $A_{1,2}$ are also
real. One should be careful with the phases. For a pure internal
interaction, the sign of the mixing matrix element $V_{12}$ is
irrelevant, it always can be changed by the redefinition of the
phase of one of the states, {\sl 1} or {\sl 2}. But with coupling
to continuum this change must be accompanied by the corresponding
phase change in the decay amplitude; therefore we cannot simply
put $A_{1,2}=\sqrt{\gamma_{1,2}}$.

Formal diagonalization of the effective Hamiltonian gives the
complex energies of the quasistationary states
\begin{widetext}
\begin{equation}
{\cal E}_{\pm}=\frac{1}{2}\left[\epsilon_{1}+\epsilon_{2}-\frac{i}{2}
(\gamma_{1}+\gamma_{2})\right]
\pm\frac{1}{2} \left \{(\epsilon_{1}-\epsilon_{2})^{2}+4v^{2}-\frac{1}{4}
(\gamma_{1}+\gamma_{2})^{2}
-i[(\epsilon_{1}-\epsilon_{2})(\gamma_{1}-
\gamma_{2})+4vA_{1}A_{2}]\right \}^{1/2}.                        \label{14}
\end{equation}
\end{widetext}
Let us check some particular cases.

(i) For the case of stable states, $A_{1,2}=0$, we come to the standard
two-level repulsion,
\begin{equation}
{\cal E}_{\pm}=E_{\pm}=\frac{1}{2}\left[\epsilon_{1}+\epsilon_{2}\pm
\sqrt{(\epsilon_{1}-\epsilon_{2})^{2}+4v^{2}}\right].       \label{15}
\end{equation}
The lower level reaches zero energy under the condition
\begin{equation}
v^{2}=\epsilon_{1}\epsilon_{2}.                               \label{16}
\end{equation}

(ii) Consider the case with no intrinsic mixing, $v=0$, and two
degenerate resonances, $\epsilon_{1}= \epsilon_{2}\equiv
\epsilon$. Then the Hamiltonian consists of the unit matrix
$\epsilon$ and the matrix $W$ of a special factorized type (rank
$r=1$) so that the correct linear combinations are the
eigenvectors of $W$; one of them, the analog of the Dicke coherent
state, should accumulate the total width, and the second should be
stable. Indeed, Eq. (\ref{14}) gives in this case
\[{\cal E}_{\pm}=\epsilon-\frac{i}{4}(\gamma_{1}+\gamma_{2})\pm
\frac{i}{4}(\gamma_{1}+\gamma_{2})\]
\begin{equation}
\Rightarrow \left \{
\begin{array}{cc}
E=\epsilon,&\Gamma=0,\\
E=\epsilon-\frac{i}{2}\Gamma, &\Gamma=\gamma_{1}+\gamma_{2}
\end{array}\right. .
                                                    \label{17}
\end{equation}
One of such situations with a bound state in the continuum was
discussed in \cite{friedrich85} and explained in terms of the
effective Hamiltonian in \cite{sokolov92}.

(iii) In contrast to the avoided crossing (\ref{15}) of stable
levels, the coincidence of two complex eigenvalues is possible. It
requires that two conditions be fulfilled,
\begin{equation}
(\epsilon_{1}-\epsilon_{2})^{2}=\gamma_{1}\gamma_{2},      \label{18}
\end{equation}
and
\begin{equation}
(\gamma_{1}-\gamma_{2})^{2}=16 v^{2}.                      \label{19}
\end{equation}
The coinciding complex energies ${\cal E}_{\pm}$, Eq. (\ref{14}),
evenly divide the trace of the Hamiltonian. 
In the energy-independent two-level 
Hamiltonian \cite{brentano99} the condition for 
crossing is
\begin{equation}
(\epsilon_1-\epsilon_2)(\gamma_1-\gamma_2)+4 v A_1 A_2=0.
\end{equation}
Which leads to crossing of either real energies $E_{+}=E_{-}$ or widths  
$\Gamma_{+}=\Gamma_{-}$ depending if the sign of 
\begin{equation}
X=(\epsilon_1-\epsilon_2)^2+4 v^2 -\frac{1}{4} (\gamma_1+\gamma_2)^2
\end{equation}
is negative or positive, respectively. Clearly, in the energy-dependent case
same remain true for the crossing of energies, other conditions, such as 
for crossing of the widths change because generally 
${\cal H}(E_{+})\ne {\cal H}(E_{-}).$   
The crossing and
anti-crossing of unstable levels were theoretically discussed also in
Ref. \cite{rotter01,magunov99} and experimentally
studied with microwave cavities \cite{philipp00}.

The secular equation for the eigenvalues can be also written in a
form explicitly separating the real, $\tilde{E}$, and imaginary,
$\tilde{\Gamma}$, parts of complex roots (here we again restore
the tilde sign in order to distinguish the roots from the running
energy value $E$). The real part of this equation gives
\begin{subequations}
\label{eq20}
\begin{equation}
\tilde{E}^{2}-\tilde{E}(\epsilon_{1}+\epsilon_{2})-\frac{\tilde{\Gamma}}
{4}(\tilde{\Gamma}-\gamma_{1}-
\gamma_{2})+\epsilon_{1}\epsilon_{2}-v^{2}=0,               \label{20}
\end{equation}
while from the imaginary part we obtain
\begin{equation}
\tilde{\Gamma}=\frac{\tilde{E}(\gamma_{1}+\gamma_{2})-\gamma_{1}
\epsilon_{2}-\gamma_{2}\epsilon_{1}+2vA_{1}A_{2}}
{2\tilde{E}-\epsilon_{1}-\epsilon_{2}}.      \label{21}
\end{equation}
\end{subequations}
The coupled Eqs. (\ref{20}) and (\ref{21}) determine $\tilde{E}$
and $\tilde{\Gamma}$. For an arbitrary energy dependence of the
amplitudes $A_{1,2}(E)$ that, in order to find the quasistationary
states, are to be taken in these equations at $E=\tilde{E}$, this
is still an implicit solution; even the number of roots can
change.

For a sufficiently strong interaction $v$, the repulsion of real
energies can bring the lower eigenvalue $E_{-}$ to zero (a
threshold value). Then both amplitudes $A_{1,2}$ disappear
together with the eigenwidth $\Gamma_{-}$, Eq. (\ref{21}). This
means that the lowest quasistationary state becomes bound under
the same condition (\ref{16}). If the mixing increases further,
the binding energy of the lower state is going down,
\begin{equation}
E_{-}\approx -\frac{v^{2}-\epsilon_{1}\epsilon_{2}}{\epsilon_{1}+
\epsilon_{2}}.                                     \label{20a}
\end{equation}
As was mentioned in \cite{brentano02} for a similar model with
only one non-vanishing $\gamma$, this is a prototype of the
dynamics leading to the binding of nuclei as $^{11}$Li where the
residual interaction among the valence neutrons is, as we have
assumed in this Section, of pairing type.

The higher level, in the point of bifurcation (\ref{16}), has the
energy
\begin{equation}
E_{+}=\frac{1}{2}\left[\epsilon_{1}+\epsilon_{2}+\sqrt{(\epsilon_{1}+
\epsilon_{2})^{2}+\Gamma_{+}(\Gamma_{+}-\gamma_{1}-\gamma_{2})}\right],
                                                                 \label{22}
\end{equation}
where $\gamma_{1,2}$ are to be taken at energy $E=E_{+}$. If the
effective Hamiltonian were energy-independent, both the real and
imaginary parts of its trace would be separately preserved by the
complex orthogonal transformation to the eigenvectors. This means
that we would always have
\begin{equation}
\Gamma_{+}+\Gamma_{-}={\rm tr}\,W=\gamma_{1}+\gamma_{2}   \label{23}
\end{equation}
and
\begin{equation}
E_{+}+E_{-}={\rm tr}\,\epsilon=\epsilon_{1}+\epsilon_{2}.  \label{24}
\end{equation}
At the bifurcation point, $E_{-}=\Gamma_{-}=0$, we would have
\begin{equation}
\Gamma_{+}=\gamma_{1}+\gamma_2, \quad  E_{+}=\epsilon_{1}+\epsilon_{2},
                                                                \label{25}
\end{equation}
while it follows from Eqs. (\ref{16}) and (\ref{21}) that
\begin{equation}
\Gamma_{+}(E_{+})=\frac{[A_{1}(E_{+})\sqrt{\epsilon_{1}}+A_{2}(E_{+})
\sqrt{\epsilon_{2}}]^{2}}{\epsilon_{1}+\epsilon_{2}}            \label{26}
\end{equation}
and
\begin{eqnarray}
\nonumber
\Gamma_{+}(E_{+})-\gamma_{1}(E_{+})-\gamma_{2}(E_{+})\\
=-\frac{[A_{1}(E_{+})
\sqrt{\epsilon_{2}}-A_{2}(E_{+})\sqrt{\epsilon_{1}}]^{2}}{\epsilon_{1}+
\epsilon_{2}}<0,                                                 \label{27}
\end{eqnarray}
in contradiction to the first part of Eq. (\ref{25}). The trace
violation occurs because the imaginary parts have their own energy
behavior with compulsory zero-energy thresholds. When the levels
are repelled by the mixing interaction, their widths are changed
by the dynamics outside the $2\times 2$ matrix. But the trace is
preserved and Eq. (\ref{25}) is fulfilled if
\begin{equation}
\frac{A_{1}(E)}{A_{2}(E)}=\sqrt{\frac{\epsilon_{1}}{\epsilon_{2}}}, \label{28}
\end{equation}
so that in the entire energy range of interest the two partial
widths grow proportionally, an interesting exceptional case.

\subsection{Scattering cross section}

In this subsection we consider the scattering cross section for
the case of two intrinsic states coupled to one open channel.
Although this cannot be observed with the scattering of a neutron
pair, the result is relevant for the excitation processes of a
Borromean system. The elastic cross section in the $s$-wave for a
relative momentum $k\propto \sqrt{E}$ is
\begin{equation}
\sigma(E)=\frac{\pi}{k^{2}}|S(E)-1|^{2},                     \label{31}
\end{equation}
where the scattering matrix is defined by Eqs. (\ref{3}) and
(\ref{4}) in terms of the effective Hamiltonian ${\cal H}$. In our
case, Eq. (\ref{13}), neglecting the potential scattering $s$, the
propagator can be easily found, and we obtain
\begin{equation}
T(E)=\frac{E(\gamma_{1}+\gamma_{2})-\gamma_{1}\epsilon_{2}-\gamma_{2}
\epsilon_{1}-2vA_{1}A_{2}}{(E-{\cal E}_{+})(E-{\cal E}_{-})},   \label{32}
\end{equation}
with the poles ${\cal E}_{\pm}=E_{\pm}-(i/2)\Gamma_{\pm}$ given by
Eq. (\ref{14}), or by a pair of coupled equations (\ref{eq20}).
One can notice that the relative sign of the matrix elements for
the direct internal interaction between the mixed states, $v$, and
for their continuum mediated interaction, $A_{1}A_{2}$, may
considerably change the resulting cross section.

In the special case [(ii), Sect. \ref{sec:5A}] of a pair of
degenerate intrinsic levels with no direct interaction, Eq.
(\ref{17}), the general result (\ref{32}) reduces formally to the
single Breit-Wigner resonance on a Dicke coherent state,
\begin{equation}
T(E)=\frac{\gamma_{1}+\gamma_{2}}{E-\epsilon+(i/2)(\gamma_{1}+\gamma_{2})}.
                                                            \label{33}
\end{equation}
The second root, $\Gamma=0$, of Eq. (\ref{17}) is decoupled from
the continuum and does not influence the scattering process. We
have to stress again that the ``widths" $\gamma_{1,2}$ in general
depend on running energy $E$.

At the bifurcation point (\ref{16}), the scattering amplitude becomes
\begin{equation}
T(E)=\frac{E(\gamma_{1}+\gamma_{2})-(A_{1}\sqrt{\epsilon_{2}}+A_{2}\sqrt
{\epsilon_{1}})^{2}}{E(E-{\cal E}_{+})},                     \label{34}
\end{equation}
where the higher root ${\cal E}_{+}$ is defined by Eqs. (\ref{22})
and (\ref{26}). At low energy, $E\rightarrow 0$, the behavior of
the scattering cross section, as well as photonuclear processes,
is determined by the actual energy dependence of decay amplitudes.

\subsection{Solutions with energy-dependent widths}
\label{ssec:c} In this and the next subsections we illustrate the
discussed above dynamics of two states coupled to a common
continuum. For all figures here we assume that $\epsilon_1= 100$
keV and $\epsilon_1= 200$ keV for the particle pair in $p$ and $s$
states, respectively. For these parameters the ground state
reaches zero energy, and thus becomes bound, at $v\approx 141$ keV
by virtue of Eq. (\ref{16}).

The picture with energy-independent widths is not consistent with
the definition of thresholds. As seen from Fig. \ref{Ligs}({\sl
a}), the residual interaction pushes the levels apart, and the
lower state crosses zero energy. However, the width $\Gamma_{-}$
of this state, dashed lines in Figs. \ref{Ligs}({\sl b}) and
\ref{Ligs}({\sl c}), is still positive. That would contradict
energy conservation. For calculations shown by solid lines in
Figs. \ref{Ligs}({\sl b}) and \ref{Ligs}({\sl c}) we account for
the squeezing of the available phase space volume that forces the
decay amplitudes to vanish once there is not enough energy for the
process to take place. Similar to Ref. \cite{brentano02}, we
assume in the low energy region the square root energy dependence
for the $s$-waves, and $\sim E^{3/2}$ for the $p$-wave,
\begin{equation}
\gamma_{2}(E)=\alpha\sqrt{E}, \quad \gamma_{1}(E)=\beta E^{3/2}. \label{35}
\end{equation}
Then the evolution of complex energies as a function of the
strength $v$ of the residual interaction is consistent with the
existence of thresholds; at $v^{2}=\epsilon_{1}\epsilon_{2}$ the
lower state becomes stationary, $E_{-}=\Gamma_{-}=0$. The
near-threshold behavior of the width is governed by the $s$-wave
component with the infinite slope, $\Gamma\sim\sqrt{E-E^{(c)}}$.
However, as $\alpha$ becomes smaller, Fig. \ref{Ligs}({\sl c}),
the singularity is getting confined to a smaller vicinity of
threshold, to the limit that at an observable scale the behavior
is dominated by the $p$-wave.

Besides the trivial situation, when the width of a particular
state vanishes due to energy conservation, blocking of the decay
via dynamical mixing at a single point corresponding to some
strength $v$ is possible. This effect of the bound state in the
continuum is seen in Fig. \ref{Ligs}({\sl b}), where a conspiracy
of the parameters leads to the vanishing width $\Gamma_{-}$ of the
lower state at energy $E_{-}$ still in the continuum, Fig.
\ref{Ligs}({\sl a}). Eqs. (\ref{20}) and (\ref{21}) with
$\Gamma_{-}=0$ show that this happens at the interaction strength
\begin{equation}
v=A_{1}A_{2}\frac{\epsilon_{1}-\epsilon_{2}}{\gamma_{1}-\gamma_{2}}.
                                                        \label{gamma0}
\end{equation}
Here $A_{1,2}$ and $\gamma_{1,2}=A^{2}_{1,2}$ are to be taken at
the energy $E_{-}$ found from Eq. (\ref{20}). For the model in
Fig. \ref{Ligs}({\sl b}) this happens at $v\approx 63$ keV. A
similar case appears in  Fig. \ref{gb005a15n} at $v\sim 185$ keV,
where both states of the system become stable only at this
particular mixing strength.

\begin{figure}
\begin{center}
\includegraphics[width=8 cm]{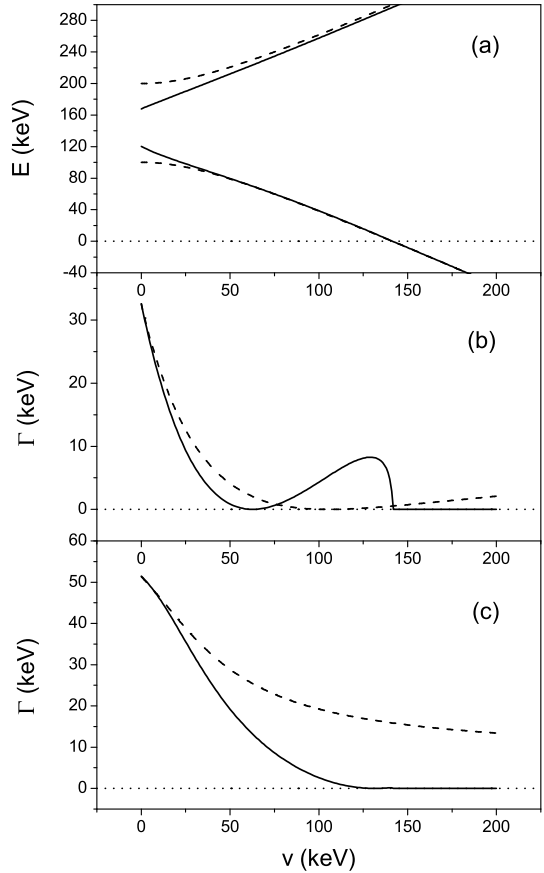}
\end{center}
\caption{Panels ({\sl b}) and ({\sl c}) demonstrate the behavior
of $\Gamma_{-}$ as a function of $v$ for energy-dependent (solid
lines) and energy-independent (dashed lines) decay amplitudes.
Selected parameters are: $A_1=8.1\,$(keV)$^{1/2}$ and $A_2=12.8\,$
(keV)$^{1/2}\,$ in the energy-independent case, and $\alpha=15$
(keV)$^{1/2}$ and $\beta=0.05$ (keV)$^{-1/2}$ in the
energy-dependent case, panel ({\sl b}); $A_1=7.1\,$(keV)$^{1/2}$
and $A_2=3.1\,$(keV)$^{1/2}$, dashed line, and
$\alpha=1$(keV)$^{1/2}$  and $\beta=0.05\,$ (keV)$^{-1/2}$, panel
({\sl c}). Parameters are selected in such a way that at $v=0$ the
two solid and dashed lines agree. The relative phases are such
that $v\ge 0$ and $A_1 A_2>0\,.$ In panel ({\sl a}) energies of
the two states are shown, solid lines, for the case relevant to
panel ({\sl b}) with the energy-dependent amplitudes, Eq.
(\ref{35}), and compared to the energies of a non-decaying system,
dashed lines. The dotted line in all three plots corresponds to
the zero value of the width or energy. \label{Ligs}}
\end{figure}

The energy dependence of the  amplitudes complicates the motion of
eigenvalues in the complex plane. Interesting features of the
level crossing are demonstrated in Figs. \ref{gb005a15n} and
\ref{gb005a10n}. It should be emphasized that, unlike in a stable
system or a system with energy-independent parameters, here the
solutions for ${\cal E}_{+}$ and ${\cal E}_{-}$ involve a
diagonalization of different matrices. The Hamiltonian matrices
differ in their imaginary part $W\,.$ The ``interaction'' between
levels occurs via common Hermitian part $H^{\circ}+V\,.$ Thus for
a general system it can be expected that bound and weakly bound
states are still strongly correlated although some new features
related to small imaginary components appear, and the usual level
repulsion is present only up to a spacing of the order of the
level width \cite{sokolov88}. For states deeply in the
continuum, however, the correlation must rely on the structure of
$W(E)\,,$ that represents features and symmetries of the
continuum.

Figures \ref{gb005a15n} and \ref{gb005a10n} also emphasize the
importance of relative phases for the matrix elements of the
internal interaction and interaction to the continuum. We can
always assume $v>0$ as was done for our figures but the change in
sign of $A_1 A_2$ leads to a system with very different
properties, compare Figs. \ref{Ligs}({\sl a}) and ({\sl b}) to
Fig. \ref{gb005a15n}.
\begin{figure}
\begin{center}
\includegraphics[width=8 cm]{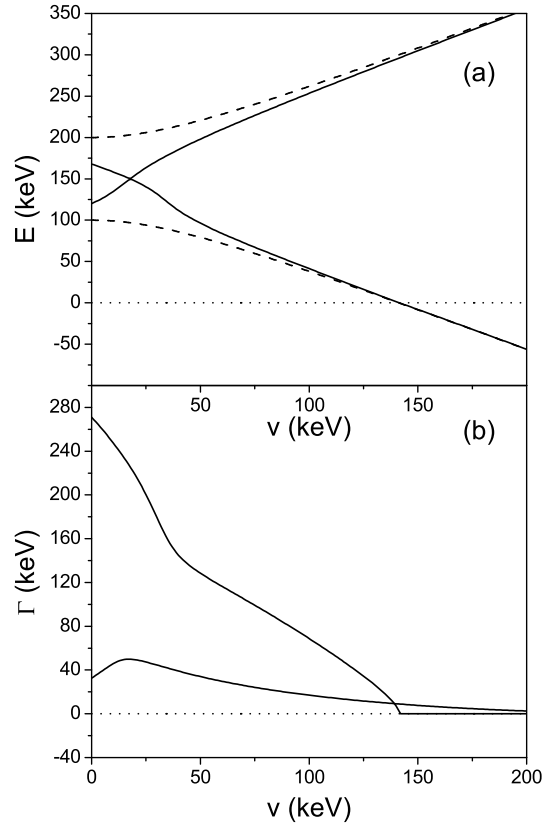}
\end{center}
\caption{\label{gb005a15n} Possible level crossing in a decaying
system. Panels ({\sl a}) and ({\sl b}) show the energies and
widths, respectively, for the two-level model of $^{11}$Li. The
upper panel, dashed lines, shows also the level repulsion in a
closed system. The parameters used are the same as in
Fig.~\ref{Ligs}({\sl b}), except for the opposite phase, $A_1
A_2<0$, for $v\ge 0\,.$}
\end{figure}
\begin{figure}
\begin{center}
\includegraphics[width=8 cm]{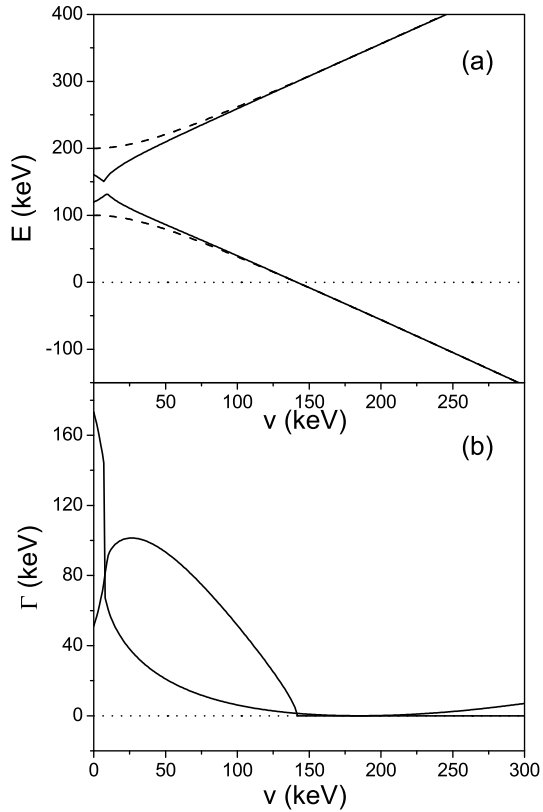}
\end{center}
\caption{\label{gb005a10n}Avoided crossing in a decaying system.
See description for Fig.~\ref{gb005a15n}. Couplings are modified to
$\alpha=10$(keV)$^{1/2}$  and $\beta=0.05\,$ (keV)$^{-1/2}\,.$}
\end{figure}

\subsection{Cross sections near threshold}
\label{ssec:d}

At the critical value of $v$, Eq. (\ref{16}), and in the low
energy region where the approximation (\ref{35}) can be valid, the
scattering amplitude (\ref{32}) is singular, $\sim E^{-1/2}$,
\begin{equation}
T(E)\approx \frac{\alpha\epsilon_{1}}{\sqrt{E}\{\epsilon_{1}+
\epsilon_{2}-(i/2)\alpha[\epsilon_{2}/(\epsilon_{1}+\epsilon_{2})]
\sqrt{E}\}}.                                           \label{36}
\end{equation}
When the interaction is over-critical, $v^{2}>\epsilon_{1}
\epsilon_{2}$, and at low energies, $E\leq |E_{-}|$ [Eq.
(\ref{20a})], we obtain
\begin{equation}
T(E)\approx \frac{\alpha\epsilon_{1}\sqrt{E}}{{\cal E}_{+}(E-E_{-})}.
                                                    \label{37}
\end{equation}
Therefore the cross section (\ref{31}) has a constant value at
threshold and behaves at low energies as $(E+|E_{-}|)^{-2}$
revealing ``attraction" to the sub-threshold region
\cite{sokolov92}. Similar near-threshold resonance phenomena were
discussed by Persson {\sl et al.} \cite{persson96}. The cross
sections shown in Fig. \ref{sigma}({\sl a}) for two over-critical
values of the interaction strength reveal a threshold behavior
characteristic for loosely bound systems that can be mistaken for
resonances. In this case we had $A_{1}A_{2}>0$ that produces only
a very broad peak (not shown on Fig. \ref{sigma}({\sl a}))
corresponding to the upper quasistationary state $E_{+}$. The next
Fig. \ref{sigma}({\sl b}) shows that in the case of $A_{1}A_{2}<0$
the interference of the internal and external interactions results
in a narrow resonance with a very high cross section at
$E=E_{+}=340$ keV, along with the maximum at zero energy (of
course, all numerical values characterize only the model
parameters).
\begin{figure}
\begin{center}
\includegraphics[width=8 cm]{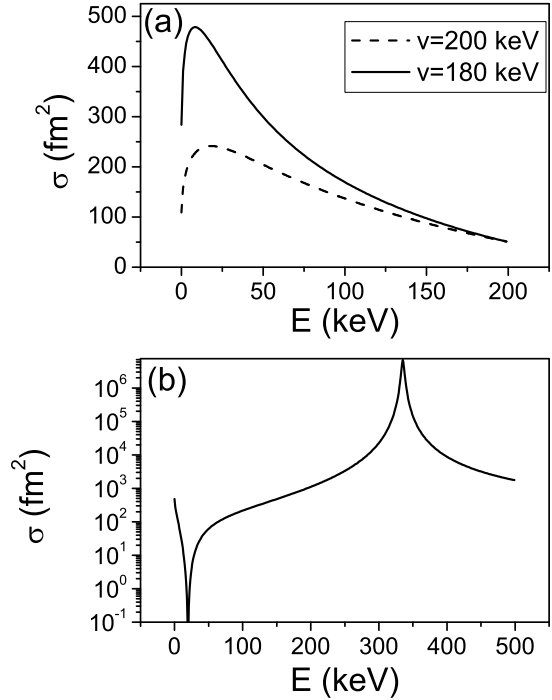}
\end{center}
\caption{The near-threshold scattering cross section is shown for
loosely bound systems with $v=180$ and $200$ keV in solid and
dashed lines, respectively. Due to similarity between the curves,
the dashed curve is not shown in panel ({\sl b}). Other selected
parameters are $\alpha=15$ (keV)$^{1/2}$  and $\beta=0.05$
(keV)$^{-1/2}\,.$ Phases are $A_1 A_2>0\,$ and $A_1 A_2<0\,$ for
panels ({\sl a}) and ({\sl b}), respectively. \label{sigma}}
\end{figure}

\section{Pairing in the continuum}

In this Section we present examples of more realistic shell model
calculations, where, in general, a large number of states is
involved and further complications arise from the conservation of
exact quantum numbers in the decay as well as from the required
self-consistency between binding energies and thresholds for one
and few-body decay channels.

The effective Hamiltonian ${\cal H}$ implicitly depends on energy
and other quantum numbers that determine if the decay is allowed
by the conservation laws and what is the near-threshold decay
rate. Below we consider examples where the intrinsic configuration
mixing is generated by the pairing interaction only. This leads to
the conservation of all partial seniorities (a number of unpaired
particles $s_j$ on each orbital $j$). We will also use $s=\sum_j
s_j$ as total seniority.

\subsection{Two-level model}

We again start with a simple model of a two-level system. We
assume here that each level can accommodate
$\Omega_1=\Omega_2=10\,$ particles and both levels can decay to a
final state that has fixed $E_f=0\,,$ their decay widths
$\gamma_1$ and $\gamma_2\,$ are different but have the same,
$\gamma_{1,\,2}(E)=\alpha_{1,\,2}\sqrt{E}$, energy dependence near
threshold. It should be noted that introduction of a synchronous
decay with $\gamma_1=\gamma_2$ does not effect the internal
dynamics since $W$ in that case is proportional to a unit matrix.
The corresponding single-particle energies are taken as
$\epsilon_1=1$ and $\epsilon_2=3\,.$ Intrinsic dynamics in this
model are generated by the constant pairing $V_{L=0}^{j j'}\equiv
G\,.$ In Fig. \ref{lscheme} the spectrum of states with seniority
$s=0$ in the system of 8 particles is shown as a function of the
pairing strength. The attractive pairing interaction pushes down
low-lying levels, forcing some of them to become bound. The ground
state becomes bound at $G\approx 0.2\,.$

Comparison of spectra with and without continuum coupling (solid
and dashed lines, respectively) shows generic features. The bound
states are not affected by the continuum coupling. The low-lying
levels, as compared to highly excited states, are less influenced
by the presence of continuum. In contrast to usual perturbation
theory, we see that the ground state and even the first excited
state once embedded in the continuum become attracted to the bulk
of other states that increases their energy. Such a situation
usually leads to an increase of the decay $Q$-value that in turn
further increases the decay width.

\begin{figure}
\begin{center}
\includegraphics[width=8 cm]{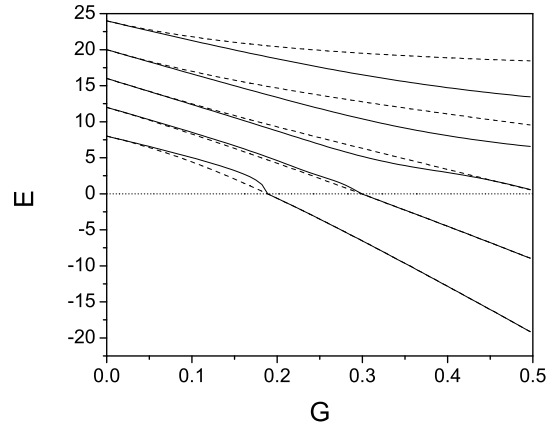}
\end{center}
\caption{The level scheme of $s=0$ states in the two-level,
8-particle system as a function of pairing strength. Solid lines
correspond to the system embedded in the continuum with the fixed
width values $\alpha_1=0.1$ and $\alpha_2=5\,.$ These curves are
compared with a non-decaying situation $\alpha_1=\alpha_2=0$ of
the usual shell-model, dashed lines. The dotted line at $E=0$
indicates the threshold location. \label{lscheme}}
\end{figure}

The following Fig. \ref{gggs}(a) demonstrates the shift $\Delta E$
of the ground state energy as a result of decay for various
choices of continuum coupling given by parameters $\alpha_1$ and
$\alpha_2\,.$ Clearly, $\Delta E=0\,$ if there is no configuration
mixing at $G=0\,,$ or once the state becomes bound. The complex
behavior of the decay width for the ground state is shown in Fig.
\ref{gggs}(b); at the critical strength the width goes to zero
with an  infinite slope, $\sim \sqrt{E}\,.$

\begin{figure}
\begin{center}
\includegraphics[width=8 cm]{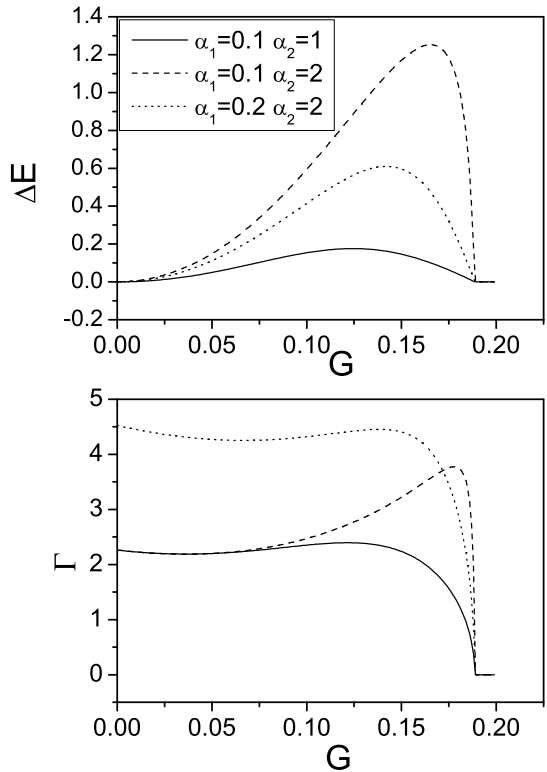}
\end{center}
\caption{The upper panel shows the shift in ground state energy
between decaying and non-decaying (usual shell-model) systems,
$\Delta E=E(\Psi)- E(\Psi_{\text{s.m.}})\,,$  as a function of
pairing strength under various assumptions for the decay rates. In
the lower panel the width of the ground state is plotted.
\label{gggs}}
\end{figure}

\subsection{Realistic pairing model}

As a demonstration of a realistic shell-model calculation we
consider oxygen isotopes in the mass region $A=16$ to 28. In this
study we use a universal $sd$-shell model description with the
semi-empirical effective interaction (USD) \cite{wildenthal84}.
The model space includes three single-particle orbitals
$1s_{1/2}$, $0d_{5/2}$ and $0d_{3/2}$ with corresponding
single-particle energies $-3.16354,\; -3.94780$ and 1.64658 MeV.
The residual interaction is defined in the most general form with
the aid of a set of 63 reduced two-body matrix elements in pair
channels with angular momentum $L$ and isospin $t$,
$\langle(j_{3}\tau_{3},j_{4}\tau_{4})Lt|V|(j_{1}\tau_{1},
j_{2}\tau_{2})Lt\rangle$, that scale with nuclear mass as
$(A/18)^{-0.3}$.

Although the full shell model treatment is possible for such light
systems, here we truncate the shell-model space to include only
seniority $s=0$ and $s=1$ states. This method, ``exact pairing +
monopole", is known \cite{volyaEP} to work well for shell model
systems involving only one type of nucleons (in the case of the
oxygen isotope chain only neutrons are involved). The two
important ingredients of nuclear forces are treated exactly by
this method: the monopole interaction that governs the binding
energy behavior throughout the mass region, and pairing that is
responsible for the emergence of the pair condensate,
renormalization of single-particle properties and collective pair
vibrations. In our exploratory study, the truncation of the large
space to the most important states is a reasonable approach since
certainly the inclusion of decay makes the computations more
numerically intense.

In the resulting shell-model description the set of the original
30 two-body matrix elements in the isospin $t=1$ channel is
reduced to 12 most important linear combinations. Six of these are
the two-body matrix elements for pair scattering in the $L=0$
channel describing pairing, and the other six correspond to the
monopole force in the particle-hole channel,
\begin{equation}
\overline{V}_{j,j'}\equiv\sum_{L\ne 0} (2L+1) \langle
(j,j')L1|V|(j,j')L1\rangle,                      \label{mono}
\end{equation}
where $j$ and $j'$ refer to one of the three single-particle levels.

We assume here that the orbital $0d_{3/2}$ belongs to the
continuum and therefore its energy has an imaginary part. In this
model we account for two possible decay channels for each initial
state $|\Phi\rangle$, a one-body channel, $c=1$, and a two-body
channel, $c=2$. The one-body decay changes the seniority of the
$0d_{3/2}$ orbital by one, from 1 to 0 in the decay of an odd-$A$
nucleus and from 0 to 1 for an even-$A$ nucleus. The two-body
decay removes two paired particles and thus does not change the
seniority. The two channels lead to the lowest energy state of
allowed seniority in the daughter nucleus, i.e. the possibility of
transition to excited pair-vibrational states is ignored. This
results in
\[e_{3/2}(\Phi)=\epsilon_{3/2}-\frac{i}{2} {\alpha_{3/2}
}\, (E_{\Phi} -E^{(1)})^{5/2}\]
\begin{equation}
 -i \,\alpha_{3/2}(E_{\Phi}-E^{(2)})^{5/2}\,,
\end{equation}
where we assumed that one and two-body decay parameters
$\gamma^{(c)}_{j}$ are related as
$\gamma^{(1)}_{3/2}=\gamma^{(2)}_{3/2}/2\equiv\gamma_{3/2}$, and
the particles are emitted in the $d$-wave with $\ell=2\,$.

These assumptions can be reviewed by examination of $^{17}$O,
where all three states with a valence particle located at one of
the single-particle orbitals can be clearly identified as the
$5/2^{+}$ ground state and $1/2^{+}$ and $3/2^{+}$ excited states.
Their energies relative to $^{16}$O exactly correspond to the
single-particle energies in the USD model. Furthermore,
experimental evidence indicates that the $3/2^{+}$ state decays
via neutron emission with the width $\Gamma(^{17}$O$)=96$ keV.
This information allows us to determine our parameter
$\alpha_{3/2}=\Gamma(^{17}$O$)/(\epsilon_{3/2})^{5/2}=0.028$
(MeV)$^{-3/2}\,.$ Other two states are particle-bound,
$\gamma_{1/2}=\gamma_{5/2}=0\,.$

Using the complex single-particle energies, the effective
non-Hermitian Hamiltonian for the many-body system is constructed
in a regular way. We treat the chain of isotopes one by one
starting from $^{16}$O. Therefore for each $A$ the properties of
the possible daughter systems $A-1$ and $A-2$ are known. Since the
effective Hamiltonian depends on energy, and all threshold
energies have to be determined self-consistently, we solve this
extremely non-linear problem iteratively. We start from the
shell-model energies $E_{{\rm s.m.}}$ corresponding to a
non-decaying system with the Hamiltonian $H$. Then the
diagonalization of ${\cal H}(E_{{\rm s.m.}})$ allows us to
determine the next approximation to the energies. This cycle is
repeated until convergence that is usually achieved in less than
ten iterations.

The results of the calculations and comparison with known
experimental data for the chain of oxygen isotopes are shown in
Table \ref{tab:oxygen}. Despite numerous oversimplifications
related to seniority truncation (some widths in the Table vanish
only due to the fact that only $s=0$ and $s=1$ states were
included), limitations on the configuration mixing and
restrictions on possible decay channels and final states, the
overall agreement observed in Table \ref{tab:oxygen} is quite
good. The results where experimental data are not available can be
considered as predictions. In our view, however, the main merit of
this calculation is in demonstrating the power of the method.

\begin{table}
$$
\begin{array}{|r|r|r|r||r|r|}
\hline
A& J & E\, {\rm (MeV) }& \Gamma \, {\rm (keV) } & E_{\rm exp.} \, {\rm (MeV) }& \Gamma_{\rm exp.} \, {\rm (keV) } \\
\hline
16& 0& {\bf 0.00} & 0 &  {\bf 0.00} & 0 \\
17& 5/2 & {\bf -3.94 } & 0 & {\bf -4.14} & 0 \\
17& 1/2& 0.78& 0 & 0.87 & 0 \\
17& 3/2& 5.59& 96 & 5.08 & 96 \\
18& 0 & {\bf -12.17}& 0 & {\bf -12.19} & 0 \\
19& 5/2 & {\bf -15.75} & 0 & {\bf  -16.14} & 0 \\
19& 1/2& 1.33& 0 & 1.47 & 0 \\
19& 3/2& 5.22& 101 & 6.12 & 110 \\
20& 0& {\bf -23.41} & 0 & {\bf -23.75} & 0 \\
21& 5/2 & {\bf -26.67} & 0 & {\bf -27.55} & 0 \\
21& 1/2& 1.38& 0 & & \\
21& 3/2& 4.60& 63 & &  \\
22& 0& {\bf -33.94 }& 0 & {\bf -34.40 } & 0 \\
23& 1/2 & {\bf -35.78} & 0 & {\bf -37.15} & 0 \\
23& 5/2 & 2.12& 0 & &  \\
23& 3/2 & 2.57& 13 & &  \\
24& 0& {\bf -40.54} & 0 &{\bf -40.85} & 0\\
25& 3/2& {\bf -39.82}& 14 & &  \\
25& 1/2& 2.37& 0 &  &  \\
25& 5/2 & 4.98& 0 & &  \\
26& 0& {\bf -42.04} & 0  & & \\
27& 3/2 & {\bf -40.29} & 339  & & \\
27& 1/2& 3.42& 59 &  &  \\
27& 5/2 & 6.45& 223 & &  \\
28& 0& {\bf -41.26} & 121 &  & \\
\hline
\end{array}
$$
\caption{Seniority $s=0$ and 1 states in oxygen isotopes. Energies
and neutron decay widths are shown. Results are compared to the
known data. Ground state energies relative to the $^{16}$O core
are given in bold. The rest of the energies are excitation
energies in a given nucleus. \label{tab:oxygen}}
\end{table}

\section{Conclusions}

The goal of the present paper is to demonstrate a variety of
results that can be obtained with the use of an effective
non-Hermitian Hamiltonian incorporated into the standard framework
of the nuclear shell model. Although those ideas are known for a
long time, right now it seems to be an appropriate moment to
revive them and convert into a working tool for the solution of
numerous practical problems of nuclear, and supposedly more
general many-body, theory. In all cases when a many-body quantum
system of strongly interacting particles is loosely bound, the
interplay of the continuum and intrinsic structure is getting
crucial, and the phenomena on the borderline between the bound
states and reaction channels become exceedingly important.
Therefore the formalism that would allow for a unified description
of interrelated structure and reaction aspects is especially
needed, and many attempts in this direction made during recent
years clearly demonstrate this need.

We illustrated the richness and nontrivial character of physics
revealed by the complicated interplay of internal and external
dynamics using a hierarchy of examples, from the simplest ones to
less obvious to realistic many-body problems. Among the most
interesting phenomena emerging here we can mention the
redistribution of the widths, similar to the Dicke superradiance
in optics, and the segregation of direct processes from those
going through the compound nucleus stage; interference of the
``normal" intrinsic residual interaction and the interaction
mediated by the excursion into open decay channels; dynamics of
the poles in the complex energy plane with unusual crossings and
anticrossings; emergence of bound states from the continuum;
typical behavior of the reaction cross sections in the presence of
loosely bound states. The correct account for the threshold
singularities of the amplitudes of the processes at low energies
was an indispensable part of the entire formalism. Finally, we
have shown how realistic problems of nuclear structure can be
solved with the aid of this method. In particular, the hybrid of
the exact solution for the pairing interaction with the
interaction through the continuum seems to be a promising
instrument for future development of theory.

Certainly, the practical implementation of the method may be more
complicated than in the standard shell model with bound states
only. The self-consistency problems of two types, namely (i) a
regular solution for the complex energies of quasistationary
states governed by the energy-dependent Hamiltonian and (ii) the
consistent determination of bound state energies, open channels
and reaction thresholds for a chain of nuclides connected by those
channels, may require new computational efforts.

The main theoretical problem that was not discussed above is
related to the residual interaction necessary for the very
formulation of the shell model problem in the presence of the
continuum. In principle, the effective interaction should be
energy-dependent and complex; it has to be consistent with the
rest of the shell model input, including the amplitudes of the
coupling to closed and open channels. This is a serious challenge
for the future that requires a new insight into the whole physics
on the borderline between structure and reactions.

\begin{acknowledgments}
The authors would like to acknowledge support from the NSF, grant
PHY-0070911, and from the U. S. Department of Energy, Nuclear
Physics Division, under contract No. W-31-109-ENG-38. V.Z. is
thankful to the participants of the Workshop on Continuum Aspects
of the Nuclear Shell Model (Trento, June 2002), where a part of
this work was presented, for interest and discussions, and to the
ECT$^\star$ for hospitality and support.
\end{acknowledgments}

\end{document}